\documentclass[sigconf]{acmart}

\AtBeginDocument{%
  }

\copyrightyear{2025}
\acmYear{2025}
\setcopyright{rightsretained}
\acmConference[CHI EA '25]{Extended Abstracts of the CHI Conference on Human Factors in Computing Systems}{April 26-May 1, 2025}{Yokohama, Japan}
\acmBooktitle{Extended Abstracts of the CHI Conference on Human Factors in Computing Systems (CHI EA '25), April 26-May 1, 2025, Yokohama, Japan}\acmDOI{10.1145/3706599.3719977}
\acmISBN{979-8-4007-1395-8/2025/04}
\begin{document}

\title[Doraemon's Gadget Lab]{Doraemon’s Gadget Lab: Unpacking Human Needs and Interaction Design in Speculative Technology}

\author{Tram Thi Minh Tran}
\email{tram.tran@sydney.edu.au}
\orcid{0000-0002-4958-2465}
\affiliation{Design Lab, Sydney School of Architecture, Design and Planning,
  \institution{The University of Sydney}
  \city{Sydney}
  \state{NSW}
  \country{Australia}
}

\renewcommand{\shortauthors}{Tran}

\begin{abstract}
Speculative technologies in science fiction have long inspired advancements in Human-Computer Interaction (HCI). \textit{Doraemon}, a Japanese manga featuring a robotic cat from the 22nd century, presents an extensive collection of futuristic gadgets—an underexplored source of speculative technologies. This study systematically analyses 379 of these gadgets, categorising them into 33 subcategories within 10 high-level groupings, to examine the fundamental human needs they address, their parallels to contemporary technologies, and their potential insights for HCI design. The findings reveal that while human needs remain constant, the ways in which technology fulfils them differ. \textit{Doraemon}’s gadgets emphasise tangible, single-purpose interactions with built-in reversibility, contrasting with the increasing complexity and software-driven nature of modern systems. By examining these speculative technologies, this study highlights alternative interaction paradigms that challenge current HCI trends and offer inspiration for future user-centred innovation.
\end{abstract}



\maketitle

\section{Introduction}

The Japanese manga\footnote{Manga refers to graphic novels or comics originating from Japan~\cite{wikipedia_manga}.} series \textit{Doraemon} is one of the most beloved and enduring works in Japanese popular culture. Created by Fujiko F. Fujio, the series follows Doraemon, a robot cat from the 22nd century, who travels back in time to assist a young boy named Nobita Nobi. Equipped with a seemingly endless array of futuristic gadgets stored in his four-dimensional pocket, Doraemon helps Nobita navigate everyday challenges and mishaps, often with unintended and humorous consequences.

Many of Doraemon's gadgets are considered \textit{`user-friendly technology at its finest.'} They are described as \textit{`portable; have no intrinsic weight; readily expand or shrink as needed; rarely break down; and easily transcend time, space, gravity, energy, and volume. They are an impressive testimony to the standards of quality control and innovation that exist in the twenty-second century'}~\cite{craig2015japan}. Although these gadgets exist within a fictional world, they reflect timeless challenges and aspirations directly relevant to modern Human-Computer Interaction (HCI). Their speculative functions offer a unique lens to examine how technology addresses human needs and how interaction design paradigms evolve over time. Therefore, this study systematically analyses 379 gadgets from Doraemon, seeking to answer the following research questions (RQ):

\begin{itemize}
    \item \textbf{RQ1}: What fundamental human desires and challenges do Doraemon’s gadgets reflect, and how do they compare to contemporary technological needs?
    \item \textbf{RQ2}: How can the speculative functions of Doraemon’s gadgets inspire interaction design paradigms in HCI?
\end{itemize}

The findings reveal that while human needs remain consistent over time, the ways in which technology addresses them differ. Doraemon envisions solutions as physical, tangible devices with intuitive, often single-purpose interactions, whereas contemporary HCI increasingly relies on software-driven, networked, and multi-functional systems. This study contributes (1) a comprehensive categorisation of Doraemon's speculative technologies, linking them to human needs and real-world counterparts, and (2) an analysis of interaction paradigms in speculative fiction, identifying design elements that may inspire future HCI research.

\section{Related Work}

\subsection{Science Fiction and HCI}
Science fiction offers a largely untapped opportunity in HCI research~\cite{jordan2021science}. Don Norman~\cite{marcus1992sci} identified three key contributions of science fiction to HCI: \textit{`prediction, social sensitivity, and scenario generation.'} Plenary panels at CHI'92~\cite{marcus1992sci} and CHI'99~\cite{marcus1999opening} featured Cyberpunk authors who \textit{`portrayed distinct visions of future advanced user interface scenarios.'} These discussions underscore how science fiction not only predicts technological possibilities but also critiques societal impacts, offering speculative insights for HCI.

Numerous studies have since explored science fiction as a lens for HCI innovation. \citet{jordan2018exploring} identified 83 publications in the CHI main track incorporating science fiction as a point of analysis. For instance, \citet{troiano206shapechanging} analysed 340 science fiction films to uncover speculative ideas about shape-changing interfaces. \citet{figueiredo2015handgestures} catalogued hand gestures inspired by science fiction, while \citet{shedroff2012make} derived design lessons from science fiction interfaces, which they catalogued online. \citet{mubin2019reflecting} examined the portrayal of 18 popular science fiction robots, and \citet{krings2023if} explored conceptualisations of programming and software development in science fiction films. Beyond films, comic books also provide rich speculative insights~\cite{johnson2011science, muscio2023ambiguous} but relatively less explored.

\subsection{Doraemon as Science Fiction}

The Doraemon series has been extensively studied for its cultural, political, and developmental impacts. Scholars have highlighted its role as a vehicle of Japanese soft power, promoting cultural values and norms to global audiences~\cite{heng2017three, harris2012ambassador, setyaningsih2023watching, craig2015japan}. Additionally, its influence on teenagers’ creativity, problem-solving, and emotional intelligence has been explored~\cite{le2024impact}. While these studies provide valuable insights into Doraemon’s significance, they overlook its massive collection of gadgets, which are highly futuristic and depict advancements not only in technology but also in the lifestyles and problem-solving approaches of the envisioned 22nd century. Some gadgets even demonstrate a remarkable foresight into technological concepts that are only now becoming reality.

One compelling example appears in Doraemon Vol. 17’s chapter \textit{Weekly Nobita} (1979)~\cite{RedditDoraemonAI2025}, where Doraemon introduces a gadget resembling a box that analyses the writing and drawing style of a manga, such as one by Tezuka-sensei, to generate a new manga in the same style. This gadget parallels Generative AI systems in 2023, which create content based on specific artistic styles. While the connection is speculative, the chapter demonstrates how fiction can explore imaginative solutions to human desires for creativity and efficiency.

By analysing Doraemon’s gadgets through the lens of HCI, this study contributes to understanding speculative technologies and their interplay with human desires and usability. Furthermore, it underscores the importance of perspectives from Asian comic books (e.g., Japanese manga/anime), complementing the insights traditionally derived from Western science fiction in computing literature.

\section{Method}

\subsection{Data Collection}

According to an analysis by Yasuyuki Yokoyama of Toyama University, there are 1,963 gadgets found in 1,344 sketches of Doraemon~\cite{japantimes2004doraemon}. However, the series' creator, Fujiko F. Fujio, originally stated that Doraemon has a total of 1,293 gadgets~\cite{wikipedia_doraemon}. The exact number varies depending on the source and method of counting. 

Given the challenge of cataloguing such an extensive and varied collection, this study used the Doraemon Fandom Wiki~\cite{doraemon_gadgets} as the primary data source, which currently lists 787 gadgets. However, to ensure alignment with the original creators' vision, the analysis focused exclusively on the 379 gadgets that originate from the original manga by Fujiko F. Fujio. Gadgets introduced in later anime adaptations, movies, or other media were excluded to maintain focus on the canonical dataset. Each gadget listed in the Doraemon Fandom Wiki was cross-referenced with its appearances in the original Doraemon manga chapters. Since the information on the wiki may be incomplete, this step involved supplementing missing details and verifying the accuracy of the provided data. 



The data collection process was supported by a Python script designed to scrape information from the Doraemon Fandom Wiki. The script, along with the resulting dataset, has been made available in the \autoref{sec:script_dataset}.


\subsection{Data Analysis}

For \textbf{RQ1}, a systematic categorisation of gadgets was conducted, organising them based on the fundamental human needs and challenges they address. To contextualise these gadgets in contemporary HCI, a multi-source search was performed to identify comparable real-world technologies. This process involved AI-powered discovery, using ChatGPT, Perplexity, and Google Search to locate modern equivalents. Additionally, community knowledge was leveraged by exploring discussions on Reddit’s r/Doraemon\footnote{\url{https://www.reddit.com/r/Doraemon/}} and r/manga\footnote{\url{https://www.reddit.com/r/manga/}} subreddits, where fans have compared Doraemon’s speculative technologies with existing or emerging real-world inventions.

For \textbf{RQ2}, a thematic analysis of the gadgets' visual and functional characteristics was conducted to examine how they communicate functionality and facilitate user interaction.  

\section{Results}

\subsection{Human Needs and Evolving Technology}

\begin{figure*} [t]
    \centering
  \includegraphics[width=0.98\textwidth]{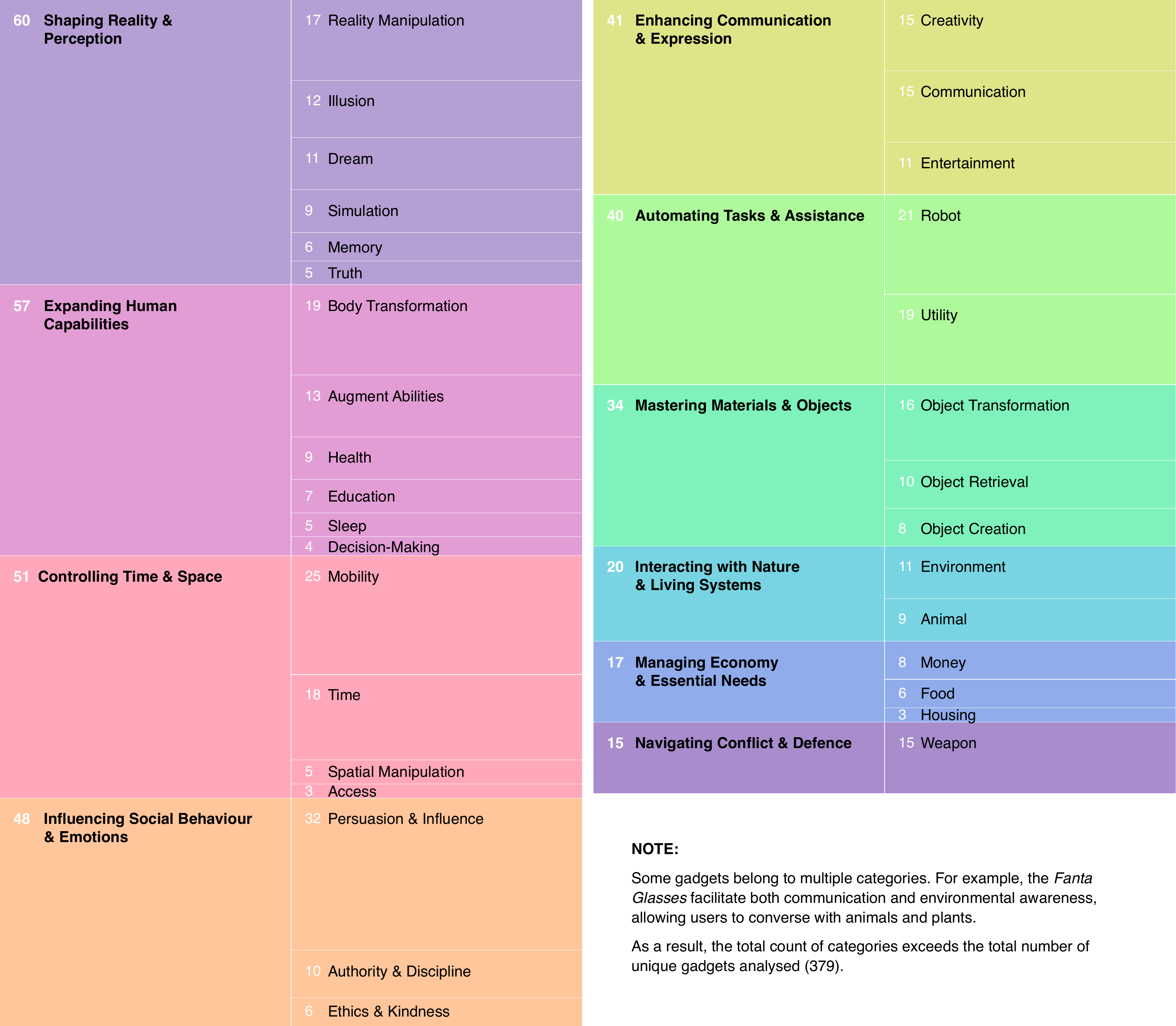}
  \caption{Icicle chart visualising the categorisation of 379 Doraemon gadgets across 10 high-level groups and 33 subcategories. The height of each section represents the number of gadgets in that category.}
  \Description{Icicle chart visualising the categorisation of 379 Doraemon gadgets across 10 high-level groups and 33 subcategories. The height of each section represents the number of gadgets in that category.}
  \label{fig:icicle_category}
\end{figure*}

The categorisation of gadgets resulted in 10 high-level groups and 33 subcategories (see \autoref{fig:icicle_category}), highlighting how speculative technology is envisioned to enhance daily life. While these gadgets reflect recurring human needs, the ways in which they address them differ from contemporary technology. The following presents a comparative overview of these speculative gadgets alongside their modern counterparts.

\subsubsection{Shaping Reality \& Perception} 
Many of Doraemon’s gadgets allow users to alter events, shape dreams, or even explore alternative realities. The \textit{What-If Phone Booth} lets users experience hypothetical scenarios by stating \textit{`What if...'} and stepping inside, instantly generating an alternative version of reality. Meanwhile, the \textit{Dream TV} allows users to observe and manipulate their own dreams, while the \textit{Forgetting Flower} can selectively erase short-term memories. 

While no real-world technology can yet rewrite reality at will, several modern equivalents hint at similar capabilities. Virtual Reality (VR) and Augmented Reality (AR) provide immersive simulations that allow users to explore alternative environments or scenarios~\cite{simeone2022immersive}, including reliving dreams~\cite{liu2024virtual}. Generative AI and deepfake technologies~\cite{shoaib2023deepfakes} enable the creation of hyper-realistic digital content, blurring the line between truth and fabrication. In neuroscience, lucid dreaming techniques~\cite{stumbrys2012induction} and memory-altering drugs~\cite{kolber2011give} are being explored to help individuals control subconscious experiences or suppress unwanted memories. Unlike these emerging technologies, Doraemon’s speculative gadgets achieve reality manipulation instantly, reversibly, and without invasive procedures.

\subsubsection{Expanding Human Capabilities} 
Many of Doraemon’s gadgets are designed to enhance human abilities, pushing the boundaries of what individuals can do. Some improve physical and cognitive skills, while others aid in decision-making, health, and well-being. The \textit{Insect Pills}, for example, grant users insect-like traits such as flight, enhanced strength, and resilience. Similarly, the \textit{Good Points Choice Board} allows users to dynamically adjust their personal attributes, such as intelligence, physical strength, or attractiveness, at the cost of reducing other traits. 

Modern technologies attempt to achieve similar enhancements. The concept of Assistive Augmentation in HCI reframes technology as an integrated extension of the human body, mind, and identity~\cite{tan2025assistive}. Exoskeletons~\cite{herr2009exoskeletons} provide mechanical augmentation for physical strength and endurance, while Brain-Computer Interfaces~\cite{wolpaw2002brain} explore ways to enhance cognitive functions and even enable direct thought-based control of devices. Advances in genetic engineering~\cite{traeger2017pre} and neurostimulation~\cite{krishna2024neuromodulation, sandoval2023advances} also hint at future possibilities for personalised human augmentation. Additionally, 3D bioprinting is being explored for the artificial creation of human skin, tissue, and internal organs~\cite{sah2019benchtop}. However, unlike Doraemon’s gadgets, which provide immediate and reversible, real-world approaches to augmentation are still constrained by ethical concerns, accessibility, and the limitations of current scientific understanding.

\subsubsection{Controlling Time \& Space}
Some gadgets challenge fundamental constraints, allowing users to travel through time, manipulate space, or move freely across dimensions. The \textit{Time Torch}, a simple flashlight-like gadget, reveals the flow of time, allowing users to visualise how their actions impact the past and future. Meanwhile, the \textit{Anywhere Door} offers instant travel to any location within 100,000 light years. 

Current technologies cannot yet match these speculative abilities, modern innovations aim to bridge similar gaps. AR and digital time-lapse visualisations provide ways to comprehend changes over time. For spatial manipulation, autonomous drone delivery~\cite{saunders2024autonomous}, high-speed transportation systems (e.g., Hyperloop)~\cite{WikipediaHyperloop}, and advancements in quantum teleportation research~\cite{WikipediaQuantumTeleportation} reflect ongoing efforts to overcome physical distance. Additionally, AR and MR environments offer users the ability to `transport' themselves to distant or even fictional locations through immersive experiences. Comparing to Doraemon’s technology offering instantaneous, infrastructure-free solutions, real-world technologies remain constrained by physics, energy demands, and economic feasibility. 

\subsubsection{Influencing Social Behaviour \& Emotions}
From persuasion and discipline to ethical enforcement, some gadgets shape human interactions and societal norms. The \textit{Go-Home Music Box} plays a tune that compels anyone who hears it to return home. 

While no real-world equivalent exists for absolute behavioural control, persuasive technologies and algorithmic nudging play increasingly significant roles in shaping human actions. Digital wellness tools~\cite{android2024wellbeing} like screen time limits and app blockers encourage self-discipline, while AI-driven social media algorithms influence opinions and behaviours through curated content~\cite{brady2023algorithm}. The key distinction between Doraemon’s speculative tools and modern implementations is autonomy and agency. While current systems rely on nudging and incentivisation, Doraemon’s gadgets enforce compliance instantly and inescapably, raising ethical questions about free will and individual rights. 

\subsubsection{Enhancing Communication \& Expression}
Many gadgets encourage creative expression, media production, and social connectivity. The \textit{Voice Candy} allows users to temporarily change their voice. \textit{Translation Tool}, which enables instant understanding between different species and languages.

Modern equivalents leverage AI and digital media to enhance self-expression and social interaction. Voice modulation software and deepfake audio tools~\cite{WikipediaAudioDeepfake} allow users to modify their voices realistically. Meanwhile, real-time translation tools such as Google Translate and AI-powered language models facilitate seamless cross-language communication. While modern technology enables more immersive and personalised communication, Doraemon’s speculative gadgets eliminate technical barriers entirely—requiring no interfaces, settings, or hardware adjustments. 

\subsubsection{Automating Tasks \& Assistance}
Robots and utility tools act as extensions of human effort, assisting with mundane tasks or even replacing human labour altogether. The \textit{Helping Hand Spray} generates autonomous gloves that continue a task even after the user stops, embodying the dream of effortless productivity.

Modern equivalents include AI-driven automation and robotics that streamline daily activities. Robotic vacuum cleaners, automated food preparation systems, and AI-powered task assistants (e.g., Amazon Alexa) serve functions similar to Doraemon’s gadgets by reducing human effort in repetitive chores. Unlike modern devices that require setup, maintenance, and user input, Doraemon’s tools respond instantly, operate autonomously, and possess intuitive adaptability—traits that remain challenging in contemporary AI and robotics. 

\subsubsection{Mastering Materials \& Objects}
From effortlessly creating objects to modifying their physical properties, Doraemon’s gadgets provide unparalleled control over material resources. The \textit{Any-Maker} lets users generate and duplicate objects instantly. Paired with the \textit{4D Pocket}, which provides limitless storage and instant access to countless gadgets.

Modern digital fabrication technologies, such as 3D printing, CNC machining, generative AI design tools, and molecular assembly research~\cite{ScienceDirectMolecularSelfAssembly}, are expanding the possibilities for customisable and efficient production. Meanwhile, the concept of instant storage and retrieval is reflected in cloud storage solutions, which allow seamless access to digital resources. Research in programmable matter~\cite{derakhshandeh2014amoebot} and autonomous fabrication~\cite{spielman2023dark, AIintheChainDarkFactories} may further bridge the gap between speculative and real-world material manipulation. Yet, Doraemon’s speculative technologies go beyond mere efficiency—they eliminate logistical concerns altogether, allowing for spontaneous materialisation, seamless storage, and effortless access to tools and physical objects. 

\subsubsection{Interacting with Nature \& Living Systems}
Beyond personal utility, some gadgets bridge the gap between humans and the environment, offering control over natural elements. The \textit{All Seasons Badge} allows users to instantly switch seasons within a 100-meter radius. Meanwhile, the \textit{Fanta Glasses} enable direct communication with animals and plants, allowing humans to understand their emotions and thoughts.

In contemporary technology, weather modification efforts, such as cloud seeding, have been explored to induce rain or mitigate extreme weather~\cite{flossmann2019review, rauber2019wintertime}. Climate engineering research, including large-scale geoengineering proposals (e.g., stratospheric aerosol injection~\cite{smith2020cost}), aims to influence global temperatures. Similarly, while AI-driven animal communication research is making strides—such as projects decoding whale songs~\cite{GoogleWhaleSongs2020} and predicting pet behaviour through machine learning~\cite{chauhan2024classifying}—true interspecies conversation, as imagined by the \textit{Fanta Glasses}, remains beyond current scientific capabilities. 

\subsubsection{Managing Economy \& Daily Needs}
Many gadgets address basic necessities, from food and shelter to money management. The \textit{Gourmet Table Cloth} produces any dish on demand, while the \textit{Apartment Pretend Tree} provides instant temporary shelter by growing into a fully furnished living space before reverting to soil overnight. 

While today’s world is far from having such instant food generators, 3D-printed meals~\cite{tran20163d, dankar20183d}, lab-grown meat~\cite{lynch2019climate}, and automated food production systems~\cite{lezoche2020agri} hint at a future where food scarcity could be alleviated through technology. Automated housing solutions, such as foldable housing~\cite{AmazonPrefabHome2024} and inflatable emergency structures~\cite{melancon2021multistable}, aim to provide rapid accommodation, particularly in disaster relief. However, unlike the \textit{Apartment Pretend Tree}, which seamlessly integrates with nature, modern housing innovations still rely on physical infrastructure and human intervention. These gadgets highlight a vision where essential needs are met through self-sustaining, instantly deployable solutions.

\subsubsection{Navigating Conflict \& Defence}
An amount of gadgets serve as tools for conflict resolution—whether through direct combat, strategic defence, or playful disruption. The \textit{Timed Stupidity Bomb}, for instance, temporarily stupefies its targets. Similarly, the \textit{Toy Soldiers} deploy autonomous miniature soldiers that provide self-defence.

Modern parallels include stun devices, such as tasers and flashbang grenades, which incapacitate threats without causing permanent harm~\cite{white2007taser, US8161883B1}. Personal safety tools like pepper spray and smart security systems also reflect real-world applications of non-lethal self-defence~\cite{edwards1997evaluation}. AI-driven security drones~\cite{FlytBase2024} and home security robots~\cite{luo2005development, song2009surveillance} offer another step toward autonomous protection, though they remain under human control rather than acting independently. 


\subsection{Unique Interaction Mechanisms}

\textit{Doraemon as a Robot Companion}: Doraemon himself is designed to address recurring challenges beyond what individual gadgets handle. His role as a companion extends to understanding and contextualising Nobita's needs and deciding which gadget to provide. Unlike individual gadgets, Doraemon proactively identifies problems and selects appropriate tools, showcasing a meta-level of problem-solving not confined to one device. His presence addresses emotional and social needs, such as companionship, guidance, and support, which gadgets alone cannot fulfil.

\textit{Object-Centric and Simple-Interaction Gadgets}:  Most gadgets in Doraemon are designed as physical objects rather than abstract digital interfaces. They often require direct interaction, such as pressing a button, wearing an accessory, or activating a switch, making their functions immediately understandable. Many gadgets have a single, clear purpose, operating independently rather than relying on complex multi-step interactions. 

\textit{Visual Design Communicating Intended Function}: Many of Doraemon’s gadgets use visual cues and familiar forms to immediately communicate their purpose. Their design often borrows from everyday objects or intuitive metaphors, making them easy to understand and operate. For example: \textit{Any-Maker}, a device that fabricates objects on demand, is designed to resemble a printer. \textit{Anywhere Door}, which enables instant travel, takes the form of an ordinary door, making its function immediately clear—simply open it, and you arrive at a new location. 

\textit{Cultural and Folklore-Inspired Design}: A number of gadgets in Doraemon draw inspiration from Japanese folklore, myths, and cultural traditions. These influences shape not only their visual appearance but also their functionality. For example, the \textit{Tanabata Rocket} reflects the Japanese tradition of writing wishes on paper strips during the Tanabata festival, transforming this custom into a physical gadget for wish fulfillment. The \textit{Floating Scarf}, which enables aerial travel, draws from the hagoromo (heavenly robe) of celestial beings in Japanese mythology. 

\textit{Reversibility and Built-in Safety Mechanisms}:  Many of Doraemon’s gadgets emphasise control and user agency, allowing characters to experiment with powerful tools without irreversible risks. Some gadgets feature explicit reversibility; e.g., the \textit{Flattening Iron} enables objects to be compressed for easy transportation or storage, with the effect reversed simply by spraying water. Other gadgets rely on temporary effects; e.g., the \textit{Adventure Tea} provides users with a thrilling experience for exactly five minutes before its effects disappear.


\section{Discussion}


\subsection{The Persistence of Human Needs in Technology Design}
The challenges addressed by Doraemon’s speculative gadgets and their modern equivalents suggest that fundamental human needs remain consistent—both attempt to solve the same core problems but with different approaches. Notably, the speculative technologies imagined in Doraemon reflect the aspirations and societal concerns of the late 1960s. Many of its gadgets address needs such as convenience, security, enhanced abilities, and social interaction, mirroring technological desires of that era. Despite the passage of time, these same needs continue to drive technological innovation today. This observation supports frameworks such as Max-Neef’s taxonomy of Fundamental Human Needs~\cite{max1991human}, which argues that basic human necessities persist across historical and technological contexts. Similarly, the Digital Society Index 2019 report~\cite{lau2019hierarchy} draws parallels between Maslow’s hierarchy of needs and the digital economy, demonstrating how technology primarily reshapes how these fundamental needs are fulfilled rather than introducing entirely new ones. These findings indicate that innovation in HCI is less about redefining human needs and more about refining and expanding the means through which they are met. 



\subsection{Rethinking Everyday Technology: From Screens to Embedded Interactions}

A striking aspect of Doraemon’s gadgets is how seamlessly they integrate into everyday life without requiring screens or digital interfaces. Unlike modern UI paradigms that evolved from the desktop metaphor~\cite{smith1985desktop}, where digital interactions mirror physical office objects, Doraemon’s world imagines technology as naturally embedded in physical objects. This suggests a potential reversal of current trends—rather than continuing to move interactions onto screens, HCI systems could explore ways to bring digital interactions back into the physical world. We have been seing examples of this reversal in Tangible Computing~\cite{ishii1997tangible} and Embodied Interaction~\cite{dourish2001action}. While purely physical interactions can be limiting, AR presents a compelling middle ground. AR-enabled technology could allow digital functionalities to be projected onto physical objects~\cite{han2023blend}, enabling interface-free computing where interactions feel as seamless as Doraemon’s gadgets. Instead of smartphone apps, an AR system could allow users to interact with the environment directly—such as looking at a store to see its reviews or pointing at a device to control it instantly. This approach preserves the simplicity of physical interaction while retaining the flexibility of software-based systems.

\subsection{Meta-Level Problem Solving: Towards Context-Aware Interfaces}
Doraemon, as a robotic companion, operates on a meta-level of problem-solving by not only providing tools but also identifying when and how they should be used. In contrast, modern super apps (e.g., WeChat)~\cite{steinberg2022media} integrate multiple functions, but users must still manually select the right service. AI assistants (e.g., Siri, Alexa) can handle some automation, but they still rely on user prompts rather than proactive anticipation. A shift towards context-aware computing~\cite{chang2013context, grubert2017towards, lindlbauer2019context} could make interfaces more proactive, reducing cognitive load for users. For instance, an AR-based operating system could detect user context and dynamically surface the most relevant applications—such as automatically opening navigation tools when stepping outside or surfacing note-taking apps during meetings. This shift would move HCI from a tool selection model to an anticipatory assistance model, making interactions smoother and more intuitive. 



\section{Conclusion and Limitations}

By analysing 379 gadgets in the Doraemon's world of speculative technology, this study explored how timeless human needs persist even as technology evolves. The gadgets—tangible, intuitive, and seamlessly embedded in daily life—offer a refreshing contrast to today’s screen-based, software-driven systems. They invite us to imagine a future where interactions feel as natural as picking up a tool, rather than navigating menus. 

The study has several limitations. First, as it focuses on speculative technologies from Doraemon, the analysis prioritises interaction paradigms and conceptual insights over technical feasibility. Consequently, gadgets may function beyond real-world constraints, limiting their direct applicability to current HCI development but still offering valuable design perspectives. Second, the identification of modern equivalents relied on AI-powered searches, publicly available sources, and community discussions, which may not comprehensively capture all emerging or niche innovations. 

\begin{acks}
The discussion on Japanese anime with Erika Wood and Xingting Wu at OzCHI 2024, along with the Reddit post `Doraemon predicted AI image generation' by u/vinnfier in r/StableDiffusion, motivated this research study. I also thank Yifan Kong for inspiring me to pursue a solo-authored paper. While working on this paper has been a valuable experience in developing independent research skills, it has also made me appreciate the collaborative teamwork I usually have with my co-authors and the way different perspectives shape the work.
\end{acks}
\bibliographystyle{ACM-Reference-Format}
\bibliography{references}

\appendix

\section{Scraping Script and Gadget Dataset}
\label{sec:script_dataset}

\subsection{Overview}
This appendix provides details on the Python scraping script used to collect data on Doraemon's gadgets and the dataset containing 379 gadgets analysed in this study. 

\subsection{Data Collection Process}
The script employs HTTP requests (via the requests library) to fetch HTML content from the website and uses HTML parsing (via the BeautifulSoup library) to extract specific information such as gadget names, appearances, functionalities, and usage histories. The data is then organised into a structured format and saved as a CSV file for further analysis. 

A total of 401 gadgets were initially obtained from the scraping process. However, 22 gadgets lacked sufficient information—either missing a Japanese name or the chapter in which they appear—making verification impossible. These gadgets were excluded from the dataset, resulting in a final count of 379 gadgets used for analysis.

The development and refinement of the Python scripts in this appendix were supported by OpenAI's ChatGPT, which provided guidance on improving code clarity and functionality. This script is intended for research purposes only and adheres to fair use policies. 

\subsection{Access to Scripts and Data}
All scripts and data outputs have been published on the Open Science Framework (OSF): \url{https://osf.io/swgc3/}.

\end{document}